\documentclass[journal]{IEEEtran}
\IEEEoverridecommandlockouts

\usepackage{comment}
\usepackage{caption}
\usepackage{stfloats} 
\usepackage{booktabs}
\usepackage{diagbox}
\usepackage{tabularx}
\usepackage{multirow}  

\usepackage{varwidth}
\usepackage{amsmath,amssymb,amsfonts,graphicx}

\usepackage{hyperref}
\hypersetup{colorlinks=true,citecolor = blue}

\usepackage{graphicx,epstopdf,caption,url}   

\usepackage{amssymb} 
\usepackage{array}    
\usepackage{arydshln} 
\usepackage{graphics}
\usepackage{color}   
\usepackage{graphicx}   
\usepackage{graphicx,epstopdf,caption,url}   
\usepackage{multirow}  

\ifCLASSINFOpdf

\else

\fi

\begin{document}
%
\title{Open Metaverse: Issues, Evolution, and Future}

\author{
    Zefeng Chen, Wensheng Gan*, Jiayi Sun, Jiayang Wu, and Philip S. Yu,~\IEEEmembership{Life Fellow,~IEEE} \\
	
	\thanks{This research was supported in part by the National Natural Science Foundation of China (Nos. 62002136 and 62272196), Natural Science Foundation of Guangdong Province (No. 2022A1515011861), the Young Scholar Program of Pazhou Lab (No. PZL2021KF0023), Engineering Research Center of Trustworthy AI, Ministry of Education (Jinan University), and Guangdong Key Laboratory for Data Security and Privacy Preserving.}

        \thanks{Zefeng Chen, Wensheng Gan, Jiayi Sun, and Jiayang Wu are with the College of Cyber Security, Jinan University, Guangzhou 510632, China; Wensheng Gan is also with Pazhou Lab, Guangzhou 510330, China.} 

        \thanks{Philip S. Yu is with the University of Illinois Chicago, Chicago, USA.}
	
	\thanks{Corresponding author: Wensheng Gan (E-mail: wsgan001@gmail.com)}
}

\maketitle

\begin{abstract}
    With the evolution of content on the web and the Internet, there is a need for cyberspace that can be used to work, live, and play in digital worlds regardless of geography. The Metaverse provides the possibility of future Internet and represents a future trend. In the future, the Metaverse will be a space where the real and the virtual are combined. In this article, we have a comprehensive survey of the compelling Metaverse. We introduce computer technology, the history of the Internet, and the promise of the Metaverse as the next generation of the Internet. In addition, we briefly introduce the related concepts of the Metaverse, including novel terms like trusted Metaverse, human-intelligence Metaverse, personalized Metaverse, AI-enabled Metaverse, Metaverse-as-a-service, etc. Moreover, we present the challenges of the Metaverse such as limited resources and ethical issues. We also present Metaverse's promising directions, including lightweight Metaverse and autonomous Metaverse. We hope this survey will provide some helpful prospects and insightful directions about the Metaverse to related developments.
    
    \textit{Impact Statement}-- The Metaverse offers many opportunities in the realms of creativity, social interaction, virtual economy, and global participation. Despite the proliferation of Metaverse-related content, the applications and directions of this innovative technology remain largely uncharted. With the intention of providing greater insight to researchers in this field, our paper explores emerging concepts in the Metaverse, including the Human-Intelligent Metaverse, Personalized Metaverse, Metaverse-as-a-service, and advanced technologies such as 5G, AI, Blockchain, and Human-Computer Interaction. We further present the potential issues surrounding the Metaverse, such as resource limitations and ethical concerns, and propose future development directions, including the Lightweight Metaverse and Autonomous Metaverse. This paper aims to provide a comprehensive review of the diverse exploration of the Metaverse, which offers a promising area for future research opportunities. The knowledge we provide in this paper will enable researchers to better understand the Metaverse, explore its limitless possibilities, and gather valuable insights into the potential of the Metaverse to transform social interaction, economy, and creativity. 
\end{abstract}



\begin{IEEEkeywords}
    Metaverse, virtual space, digital world, Metaverse-as-a-service, evolution.
\end{IEEEkeywords}

\IEEEpeerreviewmaketitle

\section{Introduction} \label{sec:introduction}

 \IEEEPARstart{W}{ith} the development of human exploration of computers, a single computer can no longer meet all of the humans' needs. Humans first attempted to connect computers via the network in order to achieve the effects of resource sharing and collaboration with others on this network, known as the Internet \cite{cohen2013internet}, which was originally used in the military. Actually, the Internet consists of various types of data, which enables the Internet of Behaviors \cite{sun2023internet}. The earliest Internet was ARPANET, which began in 1969, with the interconnection of two network nodes between the University of California, Los Angeles (UCLA) and SRI International \cite{kleinrock2008history}. In this context, with the upgrading of technology and the improvement of demand, the speed of Internet update iterations is very fast. As of 2021, the world's population has grown to 7.9 billion, and a considerable portion (about 63\%) of the population are active Internet users. Developed nations boast an even more impressive statistic, with about 90\% of their populations accessing the Internet regularly. The Internet has enabled greater flexibility in when and where humans work, especially with the proliferation of high-speed and unmetered connections. There are many ways to access the Internet from almost anywhere. In the 2020s, with the further increase in the demand for e-commerce, remote collaboration, social networking, and entertainment on the Internet, the new generation of the Internet with immersive, real-time, and digital technologies is the future development direction \cite{messinger2008typology}. In 2021,  \textit{Microsoft}\footnote{https://www.microsoft.com/} changed its name to \textit{Meta}, and then the Metaverse \cite{sun2022metaverse, mystakidis2022metaverse} becomes popular and begins to enter the vision field of researchers from all walks of life \cite{lee2021metaverse}. Its concept meets the needs of humans on the Internet and is regarded as the prototype of the next generation Internet \cite{ramesh2022metaverse}.

The Metaverse is a shared virtual space that is collective. It describes a merged environment where tangible digital space exists alongside an augmented version of the physical world, resulting in an enriched interactive experience \cite{lee2021all}. It is a new concept that combines various technologies. In other words, with the development of data science \cite{gan2019survey,gan2021survey,fournier2022pattern}, big data \cite{gan2017data,sun2022big}, artificial intelligence (AI) \cite{huynh2023artificial}, 5G \cite{cheng2022will}, blockchain \cite{gadekallu2022blockchain}, Internet of Things (IoT) \cite{li2022internet}, interactive technologies including augmented reality (AR), virtual reality (VR), mixed reality (MR), and extended reality (XR) \cite{kozinets2023immersive}, Web \cite{gan2023web,wan2023web3}, and other technologies, the Metaverse is endowed with richer meaning and vitality. The evolution process and future direction of the Metaverse are also influenced by the development of existing technologies and the emergence of new technologies. These technologies generate huge amounts of data, much larger and more varied than the traditional Internet. As a vast space, the Metaverse also presents several challenges while providing enormous potential and possibilities for methods and solutions. As the prototype of the new generation of the Internet, the Metaverse has a wide range of application scenarios, including games \cite{wiederhold2022metaverse}, social \cite{falchuk2018social}, medical \cite{yang2022metaverse}, multi-user collaboration \cite{alpala2022smart}, education \cite{lin2022metaverse}, culture \cite{kim2021study}, etc. Therefore, the Metaverse is a topic worth discussing and attracting attention. It has great value and is on the cutting edge of technology, which could be the prototype of a new generation of the Internet. However, the Metaverse is still in the early stages of development. For many researchers and entrepreneurs interested in the Metaverse, there are many possibilities for its development but plenty of concerns and uncertainty.

To provide more advances and ideas for related studies of the Metaverse, this survey topic focuses on the issues about the Metaverse and then summarizes the various emerging concepts about the Metaverse including technology-related, human-related, and other concepts like security, intelligence, service, etc. Based on these issues, the Metaverse reveals itself to be an extraordinary network and even a world of opportunities combining biology and informatics. These exciting opportunities are going to come with many challenges. In this article, we also expound on the possible challenges and problems of the future Metaverse and propose some promising directions. We are convinced that overcoming these challenges would build the Metaverse into the beautiful, immersive, high-experience, and creative digital world of the next generation.

To summarize, the main contributions of this article are listed as follows:

\begin{itemize}
    \item  We provide comprehensive concepts of the Metaverse and the basic concepts associated with them, and we introduce the application of the Metaverse to the scenarios under these concepts.

    \item  Moreover, we expound on the challenges of the coming Metaverse, including limited resources, ethical issues, and other challenges.

    \item Furthermore, we put forward several promising directions for its development, such as lightweight Metaverse and autonomous Metaverse.

    \item  Finally, we give a conclusion to this survey and set out our future vision for the open Metaverse.
\end{itemize}

\begin{figure}[ht]
    \centering	
    \includegraphics[clip,scale=0.42]{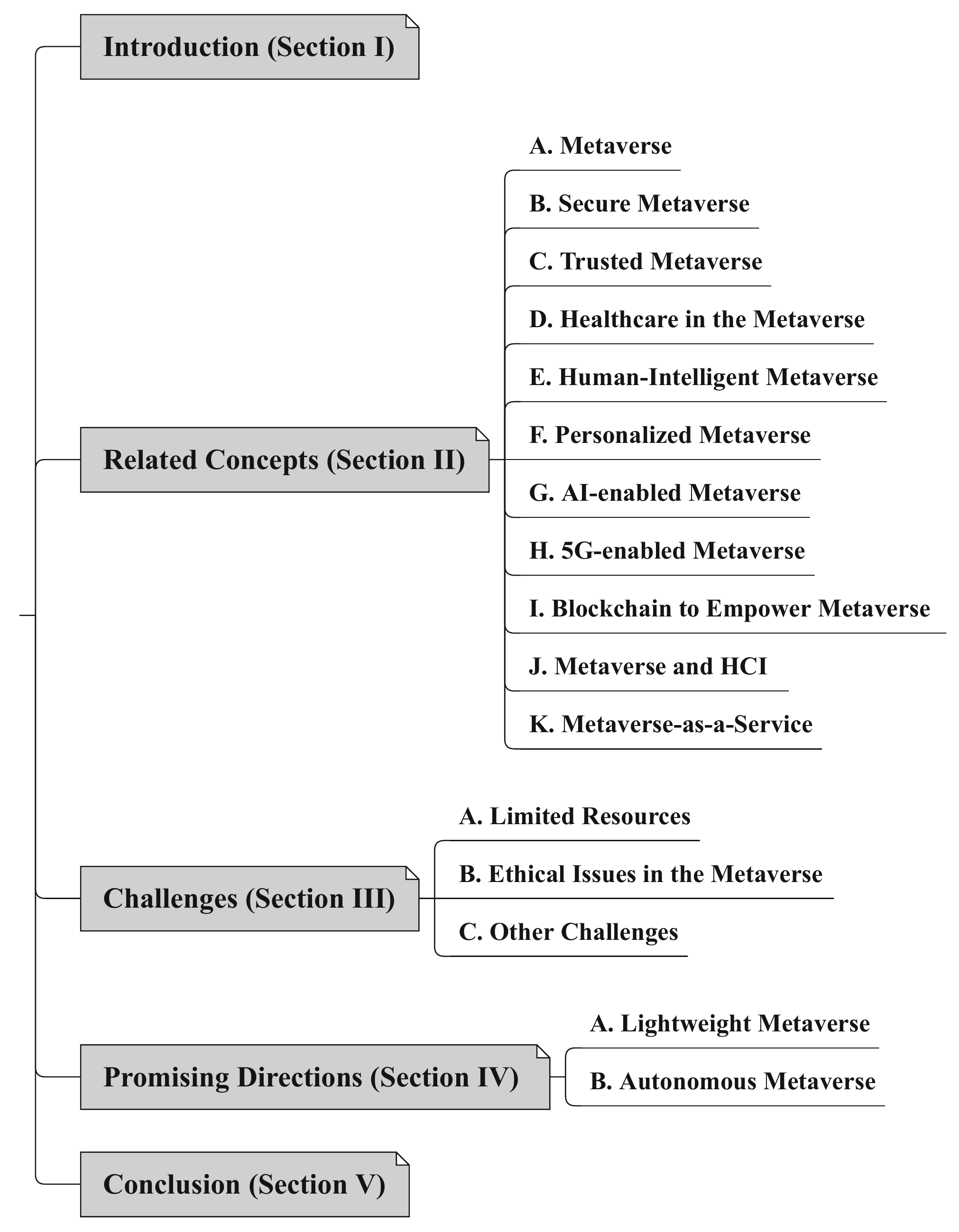}
    \caption{The outline of this article.}
    \label{fig:outline}
\end{figure}

The rest of this article is organized as follows. In Section \ref{sec:relatedwork}, we discuss related concepts of the Metaverse. We highlight the challenges in Section \ref{sec:challenges} and present several promising directions for the Metaverse in Section \ref{sec:directions}. Finally, we conclude this article in Section \ref{sec:conclusion}. The organization of this article is listed in Fig. \ref{fig:outline}. The abbreviations and their meanings are shown in Table \ref{table:abbreviations}.

\section{Related Concepts} \label{sec:relatedwork}

\subsection{Metaverse}

The concept of ``Metaverse" originated from the American fiction novel named \textit{Snow Crash} published in 1992 by Neal Stephenson \cite{lee2021all, joshua2017information}, which builds a virtual space that maps to and parallels the real world. The term ``Metaverse" consists of ``Meta" and ``Verse". ``Meta" means transcending, and ``Verse" comes from the root of the universe and represents the meaning of the universe. The combination of these two represents the meaning of ``transcending the universe" \cite{fang2021metahuman}. In addition, the future inhabitants of the Metaverse are bound to be diversified, and each individual will have a complex identity instead of a single identity. For example, in the novel \textit{Snow Crash}, each user has \textit{Avatar} in the Metaverse. 

\begin{table}[t]
	\caption{Abbreviations in alphabetical order.}
	\label{table:abbreviations}  
	\centering
	\begin{tabular}{|c|l|}
		\hline
	    \textbf{Acronym} & \qquad\qquad\quad\textbf{Explanation}  \\ \hline \hline
        5G & 5th Generation Mobile Networks \\
	    \hline	
        6G & 6th Generation Mobile Networks \\
        \hline
        AI & Artificial Intelligence \\
        \hline
        AIGC & Artificial Intelligence-Generated Content \\
        \hline
        AR & Augmented Reality \\
        \hline
        CUI & Command User Interface \\
        \hline
        DAO & Decentralized Autonomous Organization \\
        \hline
        DT & Digital Twin\\
        \hline
        GMM & Game-Mod Metaverse \\
        \hline
        GUI & Graphical User Interface \\
        \hline
        HCI & Human-Computer Interaction \\
        \hline
        HIM & Human-Intelligent Metaverse \\
        \hline
        IoT & Internet of Things \\
        \hline
        MR & Mixed Reality \\
        \hline
        NFT & Non-Fungible Token \\
        \hline
        NLG & Natural Language Generation\\
        \hline
        NLP & Natural Language Processing\\
        \hline
        NPC & Non-Player Character \\
        \hline
        NUI & Natural User Interface \\
        \hline
        PC & Personal Computer \\
        \hline
        UGC & User-Generated Content \\
        \hline
        VR & Virtual Reality \\
        \hline
        XR & Extended Reality \\
        \hline
	\end{tabular}
\end{table}

The new features of the Metaverse include a 3D virtual space that features physical persistence and convergence, combined with virtually-enhanced physical reality \cite{visconti2022physical}. It relies on the upcoming version of the Internet and features properties such as shared experiences, interconnectedness, and sensory perception\footnote{https://en.wikipedia.org/wiki/Metaverse}. Compared with the traditional Internet, the simulated real world created by the Metaverse provides users with a more real-time and immersive experience. Further, in the future, the Metaverse may not only be a concept of the Internet. As the Metaverse evolves, it's possible for human beings to cultivate novel social ties and build intentional connections with the virtual humans they create. These relationships could serve as a launching point for a new post-human society on the uncharted virtual continent, as the boundaries of the Metaverse continue to expand \cite{clark2011second}. 

Many experts believe that a true Metaverse should not just be a three-dimensional extension of the Internet, but should pursue more open standards. The Metaverse can be understood as a vast network of exponentially connected individual 3D spaces that anyone can manipulate, like a virtual social platform from which users can access anything they want, such as chatting, playing games, shopping, and watching movies. At the same time, this virtual platform should be absolutely open and fair, so it can satisfy the interests of consumers and creators, and can also foster healthy competition among developers. This means that, unlike any other game, it will be social, personal, easily accessible, and have unlimited content on an immersive, interactive, and engaging basis. It may be too early to talk about this, but it will shape how we see the world and be a major issue in our future lives.

\subsection{Secure Metaverse} 

The Metaverse is a virtual space, and how to protect the Metaverse is the most important thing for maintaining the orderly operation of the Metaverse \cite{ning2021survey}. There are two kinds of important issues in the secure Metaverse: security issues and privacy issues. The Metaverse is a world that combines technologies such as 5G, IoT, AR, VR, XR, MR, and so on \cite{mozumder2022overview}. The Metaverse will collect vast amounts of personal or public data from various data interfaces and ports \cite{sun2022big}. In other words, the Metaverse is a virtual world that is primarily made up of rich privacy or public data. Therefore, ensuring its security and privacy comes down to addressing issues at different stages of the data life cycle. These stages are data collection, data transmission, data processing, data storage, data access, and data mining \cite{gan2018privacy}. It's essential to tackle security and privacy concerns at each of these stages to ensure a secure Metaverse.

From another aspect, the issues of a secure Metaverse are supported by the Metaverse's related technologies \cite{chen2022metaverse}. For example, the current immersive experience of the Metaverse is mainly realized by interactive technologies such as AR and VR \cite{dincelli2022immersive}, and it is easier for criminals to obtain facial structures through wearable devices to crack more passwords or personal information \cite{jain2004introduction, galterio2018review}. In addition, interactive devices or interfaces can ingest data through eye tracking \cite{clay2019eye} and face/hands tracking \cite{varona2005hands}, which also poses a certain threat to the security and privacy of personal data. Therefore, the issues of a secure Metaverse are based on the security of interactive technology. In other words, the security and privacy issues of AR/VR technology lay the foundation for the security of the Metaverse to a certain extent. To give another example, there are some security technologies at present, such as blockchain \cite{zhang2019security,chen2021secure} and quantum encryption \cite{alagic2016computational,chen2021construction}. In terms of blockchain, 5G can greatly improve the transaction speed and stability of the blockchain system, and enhance the security of Metaverse financial transactions \cite{chaer2019blockchain}. The terminals of the Internet of Everything can bring more on-chain data to the blockchain. In terms of quantum encryption, the combination with 5G can improve the security and privacy of point-to-point communication in the Metaverse. In addition, 5G itself also adopts comprehensive security technology \cite{ahmad2019security}, which implements encryption and integrity protection from the two dimensions of users and information and becomes a secure information channel in the Metaverse. As one of the related technologies of the Metaverse, 5G also provides protection and escort for the secure Metaverse. Likewise, the issues of a secure Metaverse are founded on the various underlying technologies that comprise the Metaverse \cite{chen2022metaverse}.

\subsection{Trusted Metaverse}

\begin{figure}[ht]
    \centering
    \includegraphics[clip,scale=0.47]{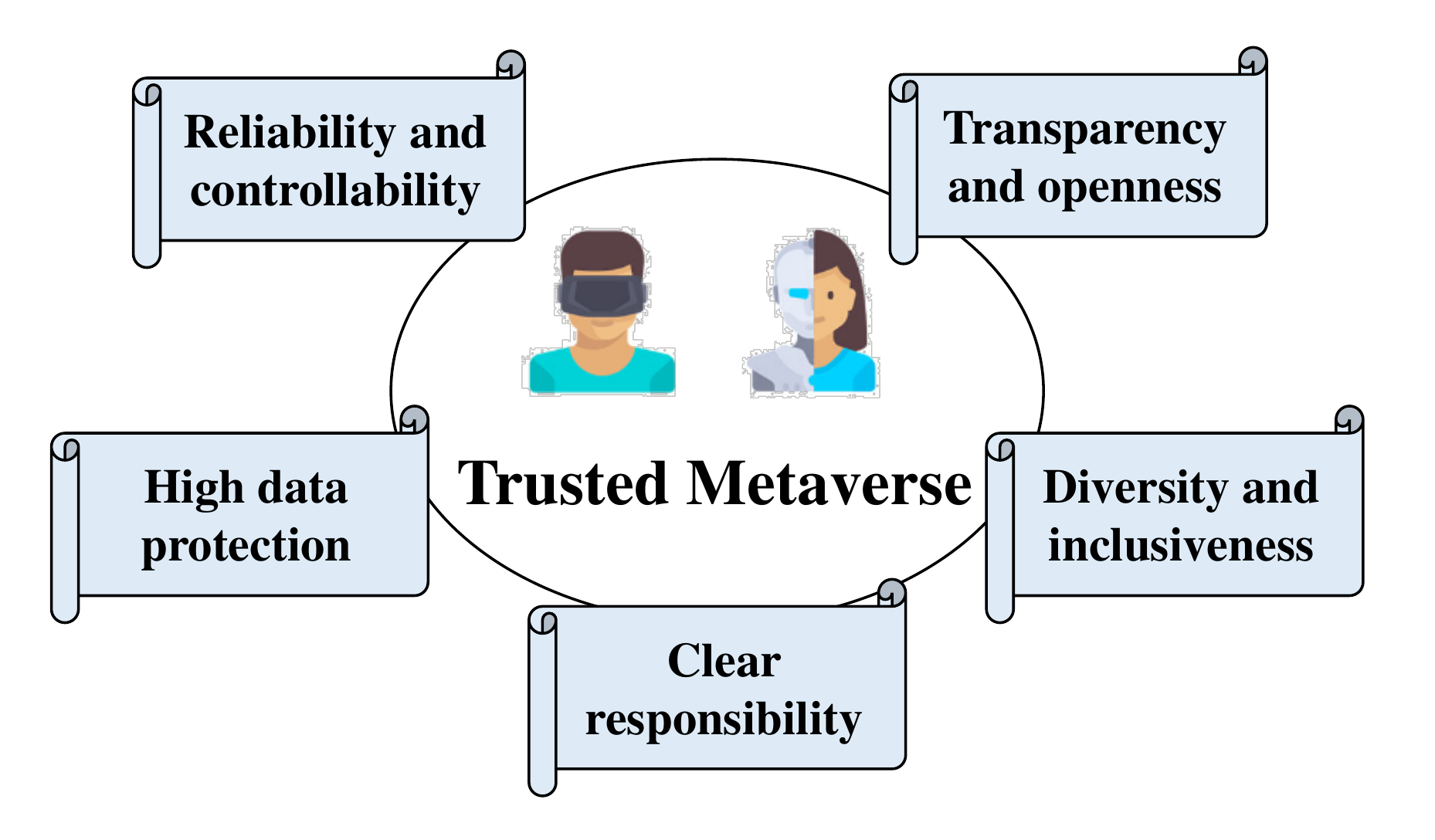}
    \caption{The features of the trusted Metaverse.}
    \label{fig:trusted}
\end{figure}

As the embryonic form of the new generation of the Internet, the Metaverse combines various technologies. In the process of the Metaverse development, some crises of trust will be exposed: data abuse leads to privacy and security risks, the complexity of decision-makers and managers poses risks, and virtual features make it difficult to define the subject of the accident, etc. Based on these crises from the perspective of technology and engineering, the trusted Metaverse implements the requirements of ethical governance and achieves an effective balance between innovative development and risk governance. Fig. \ref{fig:trusted} lists five key features of the trusted Metaverse, which are discussed clearly as follows:

\textbf{\textit{Reliability and controllability:}} The Metaverse involves many emerging technologies that are still evolving and not yet fully understood. To establish a secure and trusted Metaverse, it's crucial to prioritize technologies that are reliable and transparent rather than relying on ``black box" technologies with uncertain outcomes \cite{qayyum2022secure}.

\textbf{\textit{High data protection:}} The Metaverse combines a variety of technologies. Its numerous data requirements and numerous data interfaces make the volume of data much larger than that of the traditional Internet. Therefore, the requirements of high data protection are a vital element of the trusted Metaverse.

\textbf{\textit{Clear responsibility:}} The Metaverse has complex data ownership and processing. The complexity makes it difficult for Metaverse users to distinguish the subject of the accident. In the trusted Metaverse, clear responsibility is also an essential part. Clear responsibility means a person or a group of individuals is assigned and held accountable for specific tasks in the Metaverse.

\textbf{\textit{Diversity and inclusiveness:}} As a new generation of the Internet, the Metaverse allows people of different races, genders, ethnic groups, and skin colors to participate in this immersive world and can empower the vitality of the Metaverse as creators or inventors. A more diverse and inclusive scientific field can burst out more sparks of innovation, and a diverse and inclusive culture can make the entire Metaverse more dynamic.

\textbf{\textit{Transparency and openness:}} When developing the Metaverse, developers of the Metaverse need to disclose data permissions and where to use them to users. The transparency and openness will make the Metaverse more trustworthy.

In the actual operation and use stage of the Metaverse, it is necessary to do a good job of explaining the Metaverse system. The technical intention of disclosing the Metaverse system is to serve users better and make them safer. Various credible risks associated with the Metaverse system should continue to be monitored.

\subsection{Healthcare in the Metaverse} 

The healthcare Metaverse is to protect the information of the real world in the digital world completely. Unlike traditional 3D healthcare images, the healthcare Metaverse can form the virtual twins of patients in the virtual world according to the real-time sequence and detailed multimedia information records, so that doctors can study and explore the causes and solutions of human diseases in the virtual world without using real organisms \cite{thomason2021metahealth,bansal2022healthcare}. With the help of healthcare imaging facilities, a multiscale, multimodal, automated, and high-throughput healthcare image can be formed, and doctors can visualize the dynamic process of disease occurrence in a multi-level and panoramic way \cite{ahad2018advancements}. 

At present, Metaverse has been put into the medical field in two ways: the first is medical training, and the second is virtual surgery. In the past, young doctors were trained to accumulate experience through animal experiments, but now, with the development of the healthcare Metaverse, doctors can practice on virtual operating tables \cite{palter2010simulation}. Mainstream medical schools have expanded healthcare training by developing AI, VR, and AR technologies. Introducing the Metaverse into the healthcare field can help medical schools further turn to scene teaching. For example, students can join the virtual environment, watch the complex operation of the 3D demonstration, and even practice the operation on a person's digital twins (DTs) \cite{de2012simulation}. In the future, the Metaverse can make more contributions to healthcare treatment, such as assisting doctors to complete emergency treatment remotely \cite{pedram2020examining}. Although the telemedicine market is still dominated by video, many start-ups in the Metaverse have begun to set foot in this field. For example, \textit{ThirdEye}, which offers AR eyewear to first responders that allows for instant video and audio communication with medical professionals from a remote location. Through the AR solution, emergency responders can directly interact with the platform of see-what-I-see and broadcast emergency situations to clinicians outside the site in real time\footnote{https://info.thirdeyegen.com/thirdeye-healthcare}.

\subsection{Human-Intelligent Metaverse}

The human-intelligent Metaverse (HIM) is gradually meeting people's needs. Because of the advancement of XR and AI technologies, HIM is becoming more and more feasible. The fusion of AI and XR has led to the emergence of MR virtual worlds. According to the research conducted in \cite{zyda2022building}, HIMs are virtual environments that can simulate and communicate with humans by receiving and interpreting biosignals from various sensors, including physical, emotional, and mental cues. In contrast, HIMs are different from Game-Mod Metaverses (GMMs), game-based virtual worlds. HIMs have the ability to comprehend human physical, emotional, and mental conditions that are detected and analyzed by biosensors. Moreover, the AI characters that engage with human users within these simulated environments will possess virtual physical, emotional, and mental states that impact how they interact with and respond to these users. Future HIMs will be at the heart of future online interactive education, entertainment, telemedicine, and many other applications.

The most important part of HIM is the embodiment of digital humans. With the individuality of digital humans, the entire Metaverse could become a whole world with human intelligence. For now, the technologies of digital humans are not mature enough. Related technologies such as ray-tracing rendering technology \cite{wald2001interactive}, knowledge graph \cite{zou2020survey}, and natural language processing (NLP) \cite{chowdhary2020natural} need to be further upgraded and improved. In this way, the avatars of the Metaverse, or digital humans, will be more humane in appearance, smoother in human-computer interaction (HCI), and more intelligent in handling business tasks. When these technologies are fully mature in the near future, the commercialization of digital humans will not be just empty talk. At that point, digital humans will be able to free people from basic jobs, which will create a wealth of new business opportunities. And in the future, digital humans can already be deployed as TV presenters, brand ambassadors, online influencers, etc. Some of them have even become celebrities and garnered a lot of attention from fans, such as the Go digital human named \textit{AlphaGo} \cite{bory2019deep}. In view of this, digital humans have great potential to meet commercial purposes and great application value, making them an indispensable part of the Metaverse.

With the Metaverse becoming more industrialized and getting smarter over time, and with related technologies getting better and more mature, HIM will soon be the perfect place for the Metaverse. Of course, with the advancement and iteration of technologies, laws, and regulations on digital humans should also be gradually improved. Only in this way can HIM be better used by humans and become a Metaverse that is beneficial to and serves human beings.

\subsection{Personalized Metaverse}

Personalized Metaverse allows Metaverse to respond to the behavior of individuals or groups at the moment, which can increase the experience of Metaverse users. In other words, Metaverse is human-centric and cyberspace to serve humans \cite{yang2023human}. Artificial intelligence-generated content (AIGC) \cite{wu2023aigc} would be one of the biggest boosts in the personalized Metaverse, which combines with user-generated content (UGC), and constitutes an important part of the personalized Metaverse \cite{tang2019review, wang2022survey}. AIGC is content generated by AI, which is characterized by automated production and high efficiency. With the maturity of natural language generation (NLG) and AI models, AIGC is attracting more and more attention \cite{cao2023comprehensive}. Now it can automatically generate text, pictures, audio, video, and even 3D models and code. AIGC will greatly promote the development of the Metaverse, where a large amount of digitally native content needs to be created with the help of AI.

In the personalized Metaverse, the experience must be driven by insight into individual behavior. As virtual worlds become ubiquitous, personalization must be delivered without data. Moreover, it is no longer enough to respond in real-time just based on the past actions of many people but to respond to the actions of individuals in the present. In a personalized Metaverse, the Metaverse needs to have the following capabilities to ensure the sustainable development and vitality of the Metaverse:

\textbf{\textit{The ability to view and act on real-time customer data.}} With the help of relevant big data collection and analysis technologies, data collection can be transformed from basic storage to actionable intelligence. In the intelligent data layer, users' activities and various statistical data can be viewed in real-time. At the level of individual users, user personas can be analyzed, including what they buy, what they are doing, where they are, what devices they use, and so on. At the level of group users, it is also possible to analyze the general preferences of users, thus making the service more popular. Through these technologies, Metaverse can respond to user activities in real-time and provide personalized services better.

\textbf{\textit{The ability to capture and maintain customer interest.}} It plays a very important role in the marketing of the Metaverse. Fine-grained and real-time segmentation and personalization will enable Metaverse to provide changing services according to subtle changes in users at any moment. Metaverse combines real-time analysis and micro-analysis platforms, which can adapt to small changes in customer preferences in real-time and at any time. This level of accurate and efficient personalization plays a significant role in making the Metaverse more personalized and effective for its users.

\textbf{\textit{The ability to be genuine, authentic, and helpful.}} An important aspect of a personalized Metaverse is the realization of humanity and empathy. For example, Metaverse provides medical, education, learning, and other platforms, providing more and better opportunities to serve the public. For example, Metaverse can provide virtual sports venues, allowing people with disabilities to experience the joy and fun of sports; Metaverse can provide virtual face-to-face meeting rooms, so that scientists in foreign countries can also achieve immersive face-to-face discussions, etc.

\textbf{\textit{The ability to turn the absence of identifiers into opportunities for deeper engagement.}} As Metaverse users' data privacy awareness increases, Metaverse needs to show and inform users how their data is being used. Personalized Metaverses make users willing to share their data in exchange for something of value, such as a more tailored experience. It is an opportunity to deepen trust built on regulations and build trust in the Metaverse. In the Metaverse, data stewards can translate legal jargon into understandable language or use video to clearly articulate the sharing of first-party data, build trust, and encourage customers to share, which will help preserve the value of the underlying data.

\subsection{AI-enabled Metaverse}

Artificial intelligence (AI) \cite{honavar2006artificial}, such as federated learning \cite{chen2022federated}, has the ability of machines to perceive, synthesize, and infer information \cite{pan2016heading}. From the point of view of intelligence, the function of AI is mainly reflected in four levels, which are understanding, reasoning, learning, and interaction, respectively \cite{huang2018artificial}. In its application, AI can provide a lot of convenience for human beings. For example, AI can recognize text, images, tables, videos, and speech, and infer information through the cognitive ability to simulate data and event associations. In general, by perceiving the external environment and imitating the way of human thinking, AI has greatly improved efficiency and helped with decision-making \cite{shabbir2018artificial}. There are two primary classifications of AI, namely weak AI and strong AI, respectively \cite{flowers2019strong}. Weak AI refers to artificial intelligence that has no independent will, such as voice assistants and face recognition, and can only focus on a certain field. Strong AI is capable of acting autonomously outside of programmed processes and can even surpass humans in some aspects, such as driverless cars and intelligent robots.

AI can power the Metaverse, and the high-quality application of the Metaverse cannot be achieved without the cooperation and empowerment of AI \cite{huynh2023artificial}. Recently, Chen \textit{et al.} \cite{chen2023federated} provided a review of federated learning for Metaverse. At present, AI is a cutting-edge technology that's driving a new wave of scientific and technological revolutions along with industrial transformation. In today's society, AI technology has a significant impact and brings significant changes to many industries and societies across the world. In the new world dimension of the Metaverse, AI can not only make the Metaverse more diverse in form and more appealing in experience but also give full play to the industrial enabling effect of the Metaverse and realize the previously unrealized creativity.

Specifically, the combination of AI and the Metaverse opens up a whole new level of fidelity. AI in the Metaverse is divided into the following five use cases\footnote{https://www.xrtoday.com/virtual-reality/artificial-intelligence-in-the-metaverse-bridging-the-virtual-and-real/}:

 \textbf{\textit{Versatile avatars' creation:}} In the Metaverse, various features affected by facial expressions, emotions, hairstyles, and aging can be drawn through AI to make avatars more versatile. For example, \textit{Ready Player Me}\footnote{https://readyplayer.me} has already incorporated the use of AI technology in creating avatars for the Metaverse.

\textbf{\textit{Digital humans:}} Digital humans, like non-player characters (NPCs) in games, can respond to human actions. In the Metaverse, digital humans can be used to respond to the words and actions of humans. For example, they can serve as automated assistants in the workplace. Digital humans are built entirely using artificial intelligence technology and are central to the Metaverse's landscape. For example, the company \textit{Unreal Engine}\footnote{https://www.unrealengine.com} has invested and developed in the direction of digital humans.

 \textbf{\textit{Multilingual accessibility:}} Artificial intelligence can do the following processing of natural language: help break down natural language, convert it into a machine-readable format, analyze the contents, come up with a response, convert the results back to the language required by the users, and send it to them. The whole process is like a real conversation and takes very little time. After being fully trained by AI, it can be converted into languages around the world, so that users around the world can access the Metaverse.

\textbf{\textit{Massive scalability of virtual worlds:}} When an AI system is given past data as input, it uses the original data to train itself and generate required information as outputs. Through continuous exposure to new inputs, receiving feedback from human interactions, and applying reinforcement learning techniques, the AI improves its output and achieves increasingly satisfactory results over time. Eventually, AI will be able to perform tasks and provide output almost as well as humans. As an illustration, companies like \textit{NVIDIA}\footnote{https://www.nvidia.com} are currently training AI to generate complete virtual worlds. The aspect of the AI being able to add new worlds without human intervention will help drive the scalability of the Metaverse.

\textbf{\textit{Intuitive interfacing:}} The Metaverse has the potential to offer a profoundly realistic sensory experience using AI technology, allowing users to experience touch and other tactile sensations in virtual environments. Additionally, AI can enable voice-operated navigation, removing the need for hand-held devices and allowing users to interact with virtual objects more naturally. It will be achieved through advanced sensor technology that can detect and anticipate the user's electrical and muscular patterns, accurately interpreting their intended movements in the Metaverse.

\subsection{5G-enabled Metaverse}

The Metaverse has a close connection between reality and the virtual world. The Metaverse, with its key traits of seamless integration of virtual and real elements, constant accessibility, and high standards for security and reliability, will heavily rely on mobile communication technology (particularly 5G) to function effectively. To support these features, the Metaverse requires communication systems capable of delivering ultra-large bandwidth and ultra-high reliability. The features of 5G, including high bandwidth, low latency, reliable connectivity, and broad coverage, are vital in enabling people to access and immerse themselves in the virtual world \cite{zhang20145g,murakami2020research}. 

\begin{figure}[t]
    \centering
    \includegraphics[clip,scale=0.36]{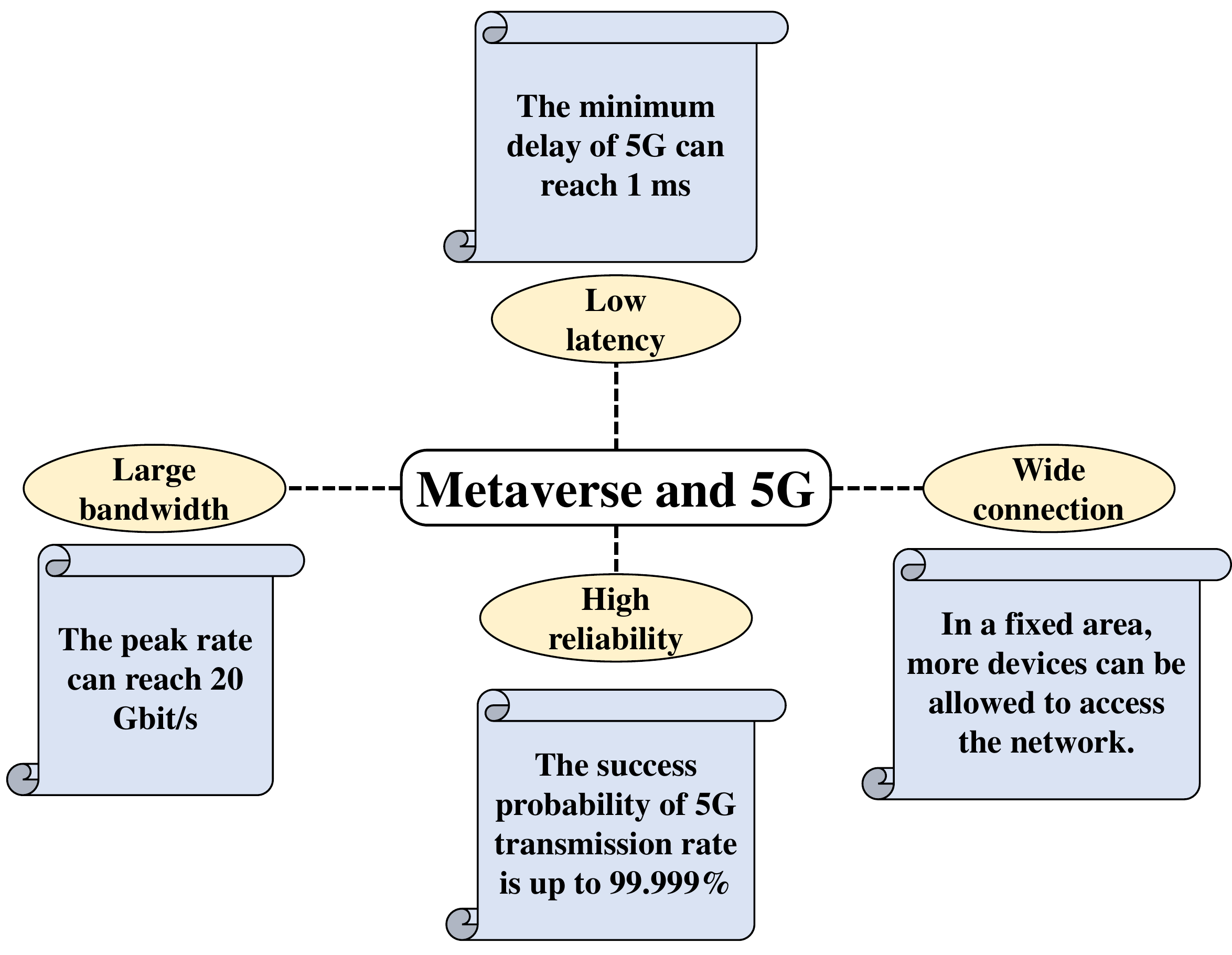}
    \caption{The features of the Metaverse and 5G.}
    \label{fig:5G}
\end{figure}

According to the characteristics of 5G, we can analyze the services that 5G can bring to the Metaverse, as shown in Fig. \ref{fig:5G}. For a large bandwidth, the rate of 5G is an order of magnitude higher than that of 4G. The peak rate can reach 20 Gbit/s, and the user's generally perceived rate can reach 100 Mbit/s \cite{nikandish2020breaking}. Moreover, for immersive 8K VR technology, the bandwidth requirement is expected to exceed 1 Gbit/s, and the peak rate of 5G can also be met. In addition, in terms of low latency, the minimum delay of 5G can reach 1 ms, and the typical end-to-end delay is 5 to 10 ms, which is far lower than human perception \cite{lema2017business}. In terms of high reliability, 5G has a reliability index requirement of more than 99.999\% in autonomous driving, remote operation, automated chemical plants, etc \cite{murakami2020research}. In the wide connection, when people are immersed in the virtual world, they need many sensors to realize virtual and real interaction \cite{khanh2022wireless}. 5G can also play a great role in this regard. The integration of 5G technology with cloud computing, cloud rendering, and other technologies will become a key measure to promote the development of the Metaverse industry. In 5G and cloud computing, connectivity, and computing power will be two important capacity resources for the development of the Metaverse \cite{skondras2019mobility}. 5G will use edge computing to improve the computing, storage capacity, and content of the cloud closer to the user so that network latency is lower and the user experience is more optimal. 5G and real-time cloud rendering can connect tens of billions or even hundreds of billions of massive sensors and truly realize the interconnection between the virtual universe and everything in the real space \cite{tang2021survey}. In the future, communication technology will also develop more and more, and 6G technology will be more suitable for the Metaverse, bringing a higher sense of experience to the Metaverse \cite{tang2022roadmap}.

\subsection{Blockchain to Empower Metaverse}

Blockchain is not a necessary technology for the Metaverse, but the characteristics of the blockchain will greatly empower the Metaverse \cite{gadekallu2022blockchain}. The unhackability, transparency, distributed ledger, and immutability of the blockchain \cite{anita2019blockchain} are key attributes for any VR technology to gain widespread adoption. There will be more hacks and data breaches in the Metaverse because the Metaverse is a completely virtual space. If users are to operate in a fully online and virtual Metaverse, one of the most basic requirements is that the underlying platform they operate on be secure, and blockchain comes in handy. Blockchain enables not only fast information confirmation but also cryptographically secure and protected transactions. Blockchain and encrypted assets are the basis of the deployment method in the Metaverse, which greatly protects the Metaverse. For example, the needs of Metaverse users are to transact and participate as easily as in person, to participate in transactions in real-time, and to trust that those transactions will be completed. Blockchain and crypto assets such as Non-Fungible Tokens (NFTs) \cite{wang2021non} can help it achieve the goal. Individuals and institutions in the Metaverse can conduct transactions in a virtual, real-time, and traceable manner through blockchain-encrypted transactions \cite{fu2022survey}.

The Metaverse is an emerging and rapidly expanding sector, and the integration of blockchain technology and cryptocurrencies will be critical to its successful development and deployment \cite{yang2018blockchain, bennett2020blockchain, raman2021world}. Blockchain technology and cryptocurrencies allow the Metaverse to enable interoperability, digital proof of ownership, a digital collection of assets (such as NFTs), value transfer through cryptocurrencies, governance, and more. The hash algorithm and timestamp technology in blockchain provide the traceability and confidentiality of the underlying data for the Metaverse. The consensus mechanism can solve the trust problem and use the distributed model to realize the self-certification of each node in the network. Most importantly, blockchain completely eliminates the intermediaries and enables direct peer-to-peer transactions, which is the core of decentralization \cite{chen2020blockchain, esmat2021novel}.

\subsection{Metaverse and Human-Computer Interaction}

\begin{figure}[t]
    \centering
    \includegraphics[clip,scale=0.45]{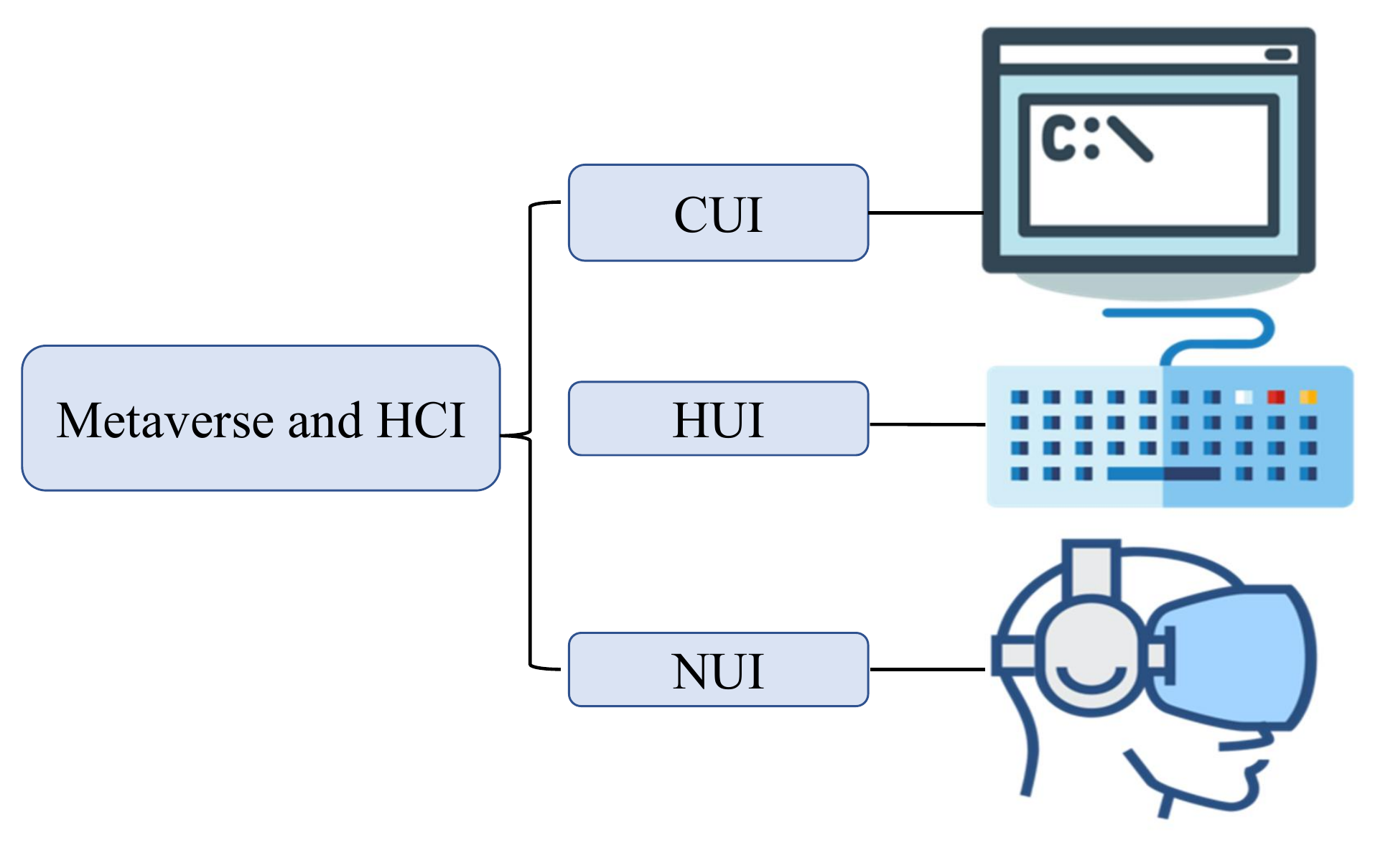}
    \caption{Metaverse and HCI.}
    \label{fig:HCI}
\end{figure}

Human-computer interaction (HCI)\footnote{https://www.interaction-design.org/literature/topics/human-computer-interaction} is also the basic capability of the Metaverse system. Its performance level directly determines the capability boundary of human beings in the Metaverse and affects the value of the Metaverse to human beings \cite{wang2022metaverse,alam2022metaverse}. In the immersive virtual world constructed by the Metaverse, how do people integrate into the virtual world? Adopting more natural and diversified HCI may be the key. There are three interaction modes, as shown in Fig. \ref{fig:HCI}. In the past, there were two common HCI modes, namely, command user interface (CUI) and graphical user interface (GUI) \cite{classen1997cui}. However, both CUI and GUI require users to learn the operations preset by the software developer. CUI obtains type instructions from the character users' interface, and GUI is to obtain the dragging instructions of the user's mouse in the graphical user interface. Neither of these two technologies can be used in the HCI technology of the Metaverse. The HCI used in the Metaverse should break through the current two technologies and adopt the most natural way of NUI \cite{joo2017study,liu2010natural}. NUI sends instructions using body posture to control the computer, such as voice, facial expression, gestures, moving the body, and rotating the head. Somatosensory interaction has become the inevitable development direction of HCI \cite{sun2022augmented}. 

However, due to the fact that the somatosensory interaction technology is not fully mature and the deep intelligent hardware industry has not yet formed a complete ecological chain, there is still a long way from having a real sense of reality in the Metaverse. The emergence of the Metaverse will force the upstream and downstream enterprises of the HCI industry chain to carry out rapid technological iteration and innovation integration, thus promoting the rapid realization of the scene. At present, some technology giants are developing intelligent glasses to realize natural interaction and only use various body postures to send commands to control the computer \cite{lee2018interaction,kim2021applications}. However, due to the loss of accurate input devices, such as the desktop mouse and mobile phone touch screen, and the lack of perceptual support for tactile feedback, the completion of these basic interactive functions is also a major challenge for these new terminal devices, such as smart glasses.

\subsection{Metaverse-as-a-Service}

The most important thing in Metaverse is service instead of sales. In other words, the group that benefits from convenience and services should be all users of the Metaverse rather than developers only. The Metaverse is committed to enabling more people to enjoy resources that were not originally involved. In consideration of the impact of economic benefits, the current Metaverse service is more about serving the masses and less about vulnerable groups. Therefore, how to spread the Metaverse to more vulnerable groups is a problem that needs to be solved in the future.

The Metaverse has revolutionary significance for the disabled \cite{seigneur2022should,duan2021metaverse}. People with disabilities frequently find it difficult to live as comfortably as non-disabled people. Some routine activities are difficult to achieve. The Metaverse can serve such people and solve their regret of not being able to participate in activities before. The social space within the Metaverse will provide advantages by replicating the real-life social experience through virtual reality and enabling individuals to communicate and engage with one another with ease, regardless of any physical limitations present in the real world \cite{oh2023social}. In the experience of leisure activities, the disabled can wear VR glasses at home to enjoy the experience of virtual cinemas, theater performances, museums, and gallery exhibitions. Moreover, the Metaverse can provide a means of traversing the digital environment for them. They can explore remote physical locations from comfortable homes through virtual reality headphones, making travel more convenient and affordable. Furthermore, Metaverse is important for children's education in impoverished areas. There has been an uneven allocation of educational resources for a long time. Children's learning is difficult to ensure quality in poor areas due to a lack of teacher allocation. The combination of Metaverse and education will become an opportunity to solve this problem \cite{lin2022metaverse,hwang2022definition,singh2022metaverse}. By integrating the Metaverse and education, it is possible to simulate a fully detailed and reproduced environment using advanced technologies such as complex lighting, mapping, and modeling. It enables students and teachers to immerse themselves in virtual reality using tools like VR glasses and other devices to participate in a range of online and offline classroom activities that cannot be replicated in traditional educational settings. By wearing VR glasses, students in poor areas can enter the virtual classroom to listen to lessons and communicate with teachers in the virtual environment. In this way, students in poor areas can also enjoy the same educational resources as urban students.

\section{Challenges}  \label{sec:challenges}

We believe the following challenging issues need to be explored and discussed. These challenges mainly include limited resources, ethical issues, and other challenges.

\subsection{Limited Resources}

The first challenge is that limited resources may cause poor service or even system collapse. We elaborate on these limited resources as follows:

\textbf{\textit{Large-scale distribution systems \rm\cite{verbraeken2020survey,sunyaev2020internet,forestiero2021agents}:}} The large-scale distribution system of the Metaverse includes the distribution of virtual goods and services in the Metaverse, which can be achieved in the following two ways: one way is that users can use virtual currency to buy and sell goods and services in the Metaverse market. Another way is to control the distribution of content within the Metaverse through the use of digital rights management systems. Overall, a large-scale distribution system within the Metaverse is critical to creating a solid and sustainable ecosystem of virtual goods and services. However, how to implement these large-scale distributed systems in the Metaverse remains an open issue.

\textbf{\textit{NLP \rm\cite{qiu2020pre,wolf2020transformers}:}} There are some NPCs that need to be built for the player in the system, and it is not easy to make these NPCs talk flexibly. Because users come from all over the world and don't speak the same language, real-time voice translation technology needs to be enhanced. Moreover, how to provide facilities for those users with visual and audio impairments also needs to be solved at present.

\textbf{\textit{How to enhance content creation:}} The current approach is to have the generated content in professional custody, generate using AI programs, provide an editor to support simulation, and support verification, profiling, and debugging at authoring time. The Metaverse is like a decentralized network that spans both the real and virtual worlds. Technically, the architecture requires more advanced user interfaces, social interactions, and a diversity of services. However, no single design or route has yet to meet all requirements. Building a Metaverse requires technologies such as synthetic visual reality, practical XR optics, real-time environmental scanning and semantic understanding, accurate physical simulation, efficient remote social interaction, and low latency for network transmission. 

In addition, mature solutions will appear and be widely used, which will set the stage for the next wave of innovation. First, standards are the foundation, which determines how strong the infrastructure of the Metaverse is. Moreover, if a platform is to be more applicable, it must meet multiple criteria. However, open standards and open resources are very important. Any company does not control a genuinely open standard. It must be multi-party and clearly defined.

\subsection{Ethical Issues in the Metaverse}
Because the Metaverse is a virtual space, there will be ethical issues that are inconsistent with or even contrary to those in the real world, which is inevitable and crucial. Therefore, dealing with the Metaverse's relevant ethical issues is a significant challenge for the Metaverse. In this section, the major ethical issues in the Metaverse are discussed.

\textbf{\textit{Pornography \rm\cite{evans2022worlds}}}: Pornography in the Metaverse is an aspect of the Metaverse that is barely discussed, which could be the biggest danger. As Esteban Hendrickx, an art director proposed that the pornography industry will undergo a revolution\footnote{https://www.intotheminds.com/blog/en/marketing-metaverse/}. It is true that the anonymity, virtuality, and immersive experience of the Metaverse can bring pleasure to human beings in terms of sex. However, there are still a lot of ethical issues worth exploring. For example, should virtual child pornography in the Metaverse be criminalized \cite{qin2022identity}? Will it lead to the proliferation of the porn industry, changing people's attitudes toward sex and harming young people if left unchecked? There are many ethical and legal issues worth discussing and waiting to be perfected.

\textbf{\textit{Attacks on the avatars}}: The Metaverse is a virtual world where real people interact through avatars. Then, some extreme methods may appear in the Metaverse, such as attacking the digital integrity of digital avatars through some means. It needs to be protected by some laws and regulations, and it is also an ethical issue worthy of discussion. Is it morally and legally permissible to kill or attack an avatar in the Metaverse? 
    
 \textbf{\textit{Death and legacy}}: The issues of death and legacy are also ethical issues worth discussing. \textit{Facebook}/\textit{Meta}\footnote{https://www.facebook.com} adds a separate tab to the profile of a deceased user, for reminiscence or mourning. While in the Metaverse, there are no vital organs in the virtual world. Will humans exist in the form of digital life? If the owner of the avatar dies, will the Metaverse house a special space for these avatars? In addition, after the death of the incarnate owner of the Metaverse, there are still some inheritance problems in the Metaverse. How will the Metaverse's virtual properties, such as NFT, be passed down to future generations? Is it necessary to collect inheritance tax?

\textbf{\textit{The form of avatars \rm\cite{cheong2022avatars}}}: In the Metaverse, it is foreseeable that future avatars may take various forms, and it is possible that two avatars may belong to the same owner. So, are all avatars acceptable in the Metaverse? In other words, do all avatars need to be in human form for sufficient immersion? Also, it is expected that some people create digital personalities by using different accounts in the Metaverse. The anthropomorphism of avatars, i.e., their appearance, would undoubtedly complicate the concept of identity in the Metaverse, would it be ethically acceptable to obscure the trajectory of one's true identity in the Metaverse?

\subsection{Other Challenges} 

In addition to the above-mentioned resource issues and ethical issues, the Metaverse has other challenges. The Metaverse industry is poised for rapid expansion and growth in multiple areas, driven by factors such as a growing user base, strong policy support, and innovation experience. Thus, as the Metaverse expands and transforms the internet ecosystem, economy, and social structures, it is expected to face new challenges as it matures. These challenges can be broadly categorized into the following areas:

\textbf{\textit{Infrastructure attack \rm\cite{jaber2022security}:}} Although the basic technology of the Metaverse provides many solutions, as a brand-new Internet prototype, the Metaverse will become very complicated in the future. The Metaverse is expected to grow into a vast, highly accessible, and dynamically optimized complex system, supported by extensive digital and traditional infrastructure. Given its deeper integration into the daily personal and professional lives of individuals than the traditional internet, any disruptions or attacks on its infrastructure could have significant economic and social consequences. In the event of an attack, invasion, or disturbance of the Metaverse's infrastructure, it could severely impact the normal functioning of the economy and disrupt social order. To prevent this from happening, sufficient measures and safeguards must be in place to protect the Metaverse's infrastructure and maintain a secure and stable environment for its users.
    
\textbf{\textit{High degree of monopoly \rm\cite{feng2021data}:}} For the reason that the Metaverse needs to realize ultra-large-scale user connection and interaction in the development stage, it is necessary to put into operation a large-scale infrastructure, so the early construction process requires capable companies to invest in huge investment and development. Therefore, there may be a possibility of being highly monopolized during the construction of the Metaverse. Moreover, when operating the Metaverse, a relatively stable service provider is also required to provide operational services, which is also a possible trigger for being monopolized. Therefore, how to prevent the emergence of monopolies in the Metaverse will be a critical issue for the future development of the industry.
    
\textbf{\textit{User's digital addiction \rm\cite{bojic2022metaverse}:}} The Metaverse will challenge the familiar physical rules of the real world and reshape production and lifestyles in the virtual world. While the immersive new visual experiences will attract more people to the Metaverse, it is important to maintain a positive relationship with the real world, manage negative effects, and address digital addiction. These challenges will need to be addressed in order to effectively leverage the potential of the Metaverse. Finding a way to effectively manage these challenges and utilize the Metaverse's potential for good will be crucial.

\section{Promising Directions}  \label{sec:directions}

The Metaverse will have a promising future. There are still many directions for the future development of the Metaverse, such as the lightweight Metaverse and the autonomous Metaverse.

\begin{table*}[t]
	\caption{The comparison of heavyweight Metaverse, hybrid lightweight Metaverse, and lightweight Metaverse.}
	\label{table:light}
	\begin{tabularx}{\textwidth}{m{3cm}<{\centering}|m{4.45cm}<{\raggedright}|m{4.45cm}<{\raggedright}|m{4.45cm}<{\raggedright}}
		\hline
		\textbf{Form} & \textbf{\qquad Heavyweight Metaverse} & \textbf{\quad Hybrid Lightweight Metaverse} & \textbf{\qquad\quad Lightweight Metaverse}\\ 
        \hline \hline
        \textbf{3D scene construction} & Developed based on traditional 3D engines such as Unreal/Unity & Developed based on traditional 3D engines such as Unreal/Unity & Anyone and any device can realize building blocks through a browser \\
        \hline
        \textbf{Rendering mode} & Local rendering & Local rendering and cloud rendering &High performance local rendering \\
        \hline
        \textbf{Creative interaction and expansion capabilities} &  Individual project team planning and development & Individual project team planning and development & One-click implementation through lightweight logic editor and seamless docking self-developed engine \\
        \hline
        \textbf{Development cycle} & 180 days or more & Around 90 days & Around 5-30 days\\
        \hline 
        \textbf{User entrance} & Independent Applications, individual engines support H5 & H5, applets, and Applications all have independent entrances & H5, applets, and Applications all have independent entrances \\
        \hline
        \textbf{Delivery effect} &  Realistic 3D effect & Texture-based 3D/partial animation-level 3D effects & Animation-level 3D effects\\
		\hline \hline
	\end{tabularx}
\end{table*}

\subsection{Lightweight Metaverse}

\begin{figure}[ht]
    \centering
    \includegraphics[clip,scale=0.44]{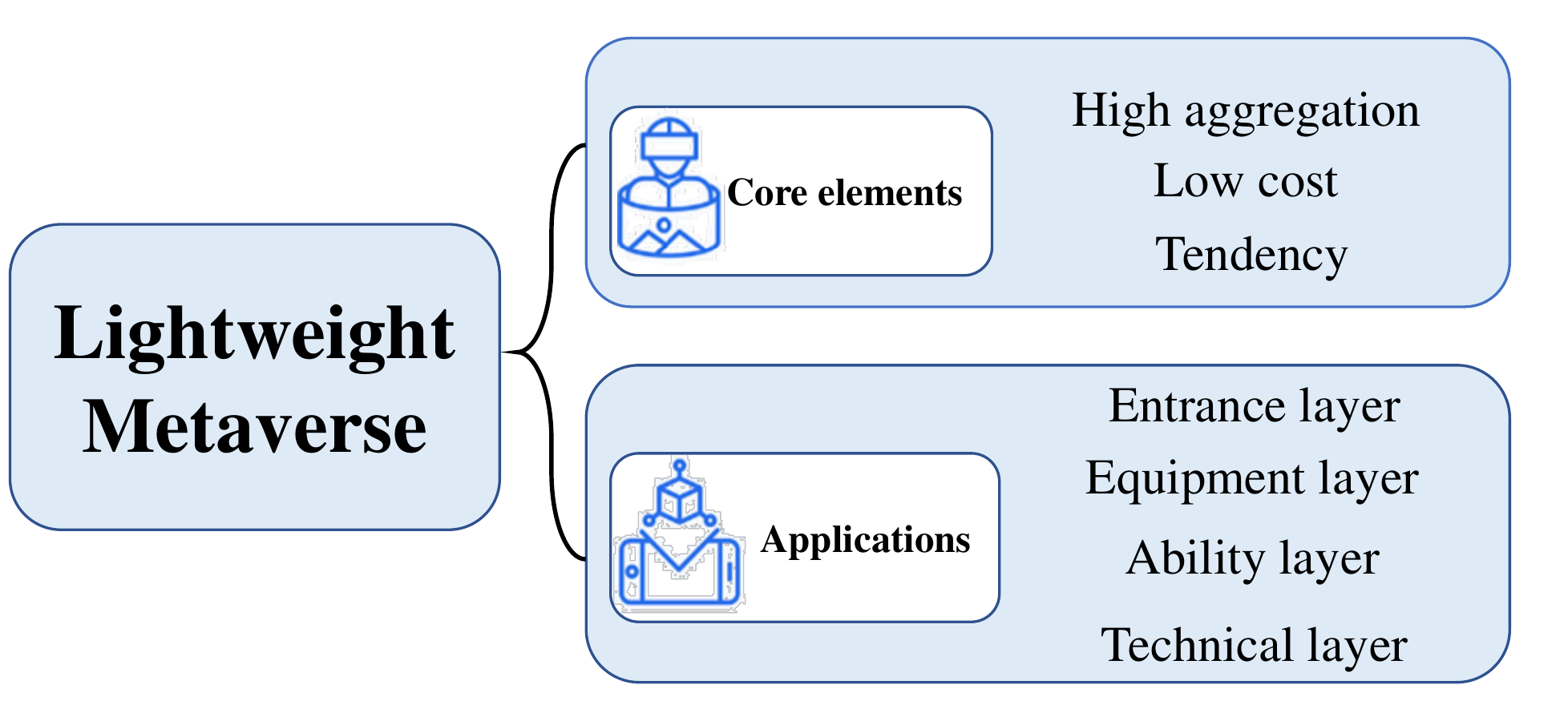}
    \caption{The core elements and applications of the lightweight Metaverse.}
    \label{fig:light}
\end{figure}

Lightweight Metaverse is a new form of decentralized, unprovoked, and lightweight Metaverse. Lightweight Metaverse is the direction for optimizing the feature that Metaverse can be accessed anytime, anywhere, and on any device. It has the advantages of low-performance requirements, a low implementation cost, and a wider application range. Lightweight Metaverse can effectively transform existing platform traffic to solve the industry's current source of Metaverse users' problems, foster platform content fission, and expand influence in the form of Metaverse. The core elements and application architecture are shown in Fig. \ref{fig:light}. 

The elements of the lightweight Metaverse are high aggregation, low cost, and tendency. High aggregation refers to being more privatized, supporting data privacy, and opening common private domain platforms such as websites with full links. Low cost is mainly manifested in the light entrance, easy operation and maintenance, and reducing the cost of communication, operation, and maintenance. This tendency mainly refers to the Metaverse model, which can resolve the pain points of traditional Metaverse products and thus quickly grow into the mainstream. The lightweight Metaverse takes unprovoked H5 and web pages as the entrance layer and supports the equipment layer of any device, such as mobile phones, PCs, and tablets. It takes digital people, NFT, and systems as the capability layer, and technologies such as engines, services, and rendering as the technical layer, thus establishing a portable and accessible Metaverse space. The lightweight Metaverse is one of the main trends of the Metaverse in the future. Its biggest advantages lie in its strong applicability, ease of large-scale promotion, and low development and operation costs.

\subsection{Autonomous Metaverse}

The digital technology of the Metaverse will let users act without middlemen and rebuild a society that lets users become active citizens. This will let them fully express their creativity and make the Metaverse more independent. Blockchain technology has the potential to power a decentralized autonomous world in the Metaverse. An autonomous model will allow us to be global citizens with digital identities and create our economies without intermediaries. Decentralization is a fascinating idea that will lead to unprecedented changes in our social and economic order, driving post-digital transformation technologies and innovations. The autonomous Metaverse will guide users to obtain direct and absolute democracy in terms of identity, ownership, and transactions, and become people who directly participate in government decision-making \cite{talmon2022foundations}.

\begin{figure}[t]
    \centering
    \includegraphics[clip,scale=0.26]{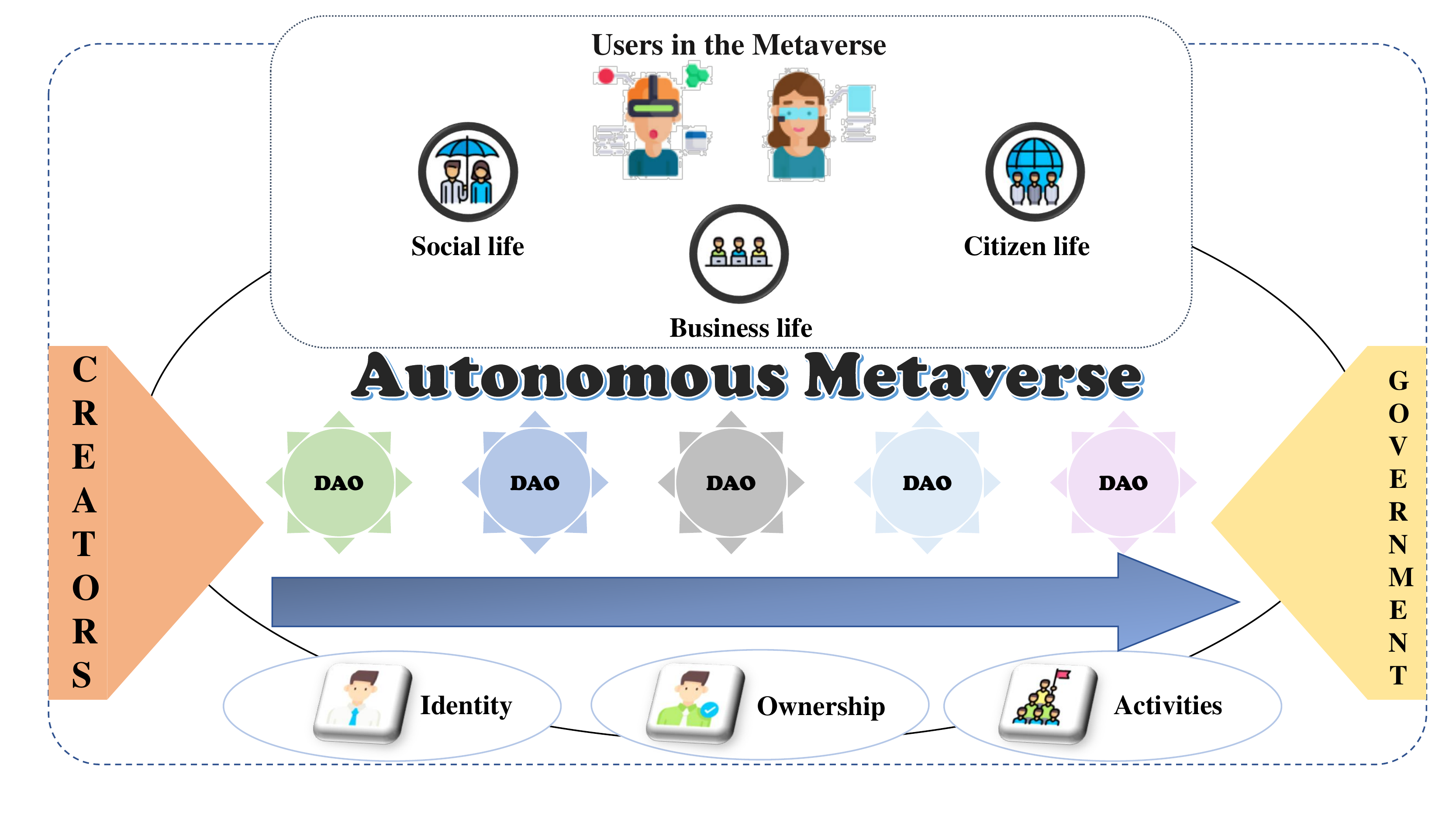}
    \caption{Schematic diagram of autonomous Metaverse.}
    \label{fig:autonomous}
\end{figure}

In our opinion, the autonomous Metaverse begins with the decentralized autonomous organization (DAO) \cite{wang2019decentralized, hassan2021decentralized}. In terms of technique, DAO encodes the operating rules of an organization into a computer program, or smart contract, to be exact \cite{frantz2016institutions, dwivedi2021formal}. The purpose of the DAO is to maintain transparency and avoid the need for a central agency intermediary. DAOs digitize users' transactions, thus playing the role of a contract. If the registry is digitized through a DAO, the management process can be radically simplified. By translating legal rules into smart contract terms, Metaverse users can be registered independently, improving efficiency and eliminating bureaucracy associated with the procedural process. More broadly, if we can combine DAOs in a way that meets the diverse needs of citizens and enterprises, there will be the birth of an autonomous Metaverse, which is a platform for cooperation and interoperability among various decentralized and autonomous organizations. With an autonomous Metaverse, our society and economy will be digitized, and the Metaverse will become a post-digital society—an interconnected, interoperable, and interdependent combination of individuals and organizations. The schematic diagram of the autonomous Metaverse is shown in Fig. \ref{fig:autonomous}.

Additionally, in the autonomous Metaverse, the role of citizens will also change. More importantly, the national borders will lose their meaning. As we move away from hierarchical structures and territorial bureaucracies, Metaverse users will transform into global citizens with a more international ethic. Apparently, in the autonomous Metaverse, citizens have more autonomy and independence.

\section{Conclusion}  \label{sec:conclusion}

The Metaverse integrates various emerging technologies of the new generation and is the embryonic form of the new generation of the Internet, which provides great possibilities for the new order of economy and society. In this article, we give a brief introduction to the Metaverse and the related concepts of the Metaverse, and then propose a series of directions for its development. However, the development of the Metaverse still faces many challenges and opportunities. Based on this, we provide some insights into the Metaverse challenges and future directions. Finally, we present the conclusion of this article. We hope that this article can provide fresh thinking directions and insights for researchers in the Metaverse, and provide useful ideas for the development of Metaverse-related academia, industry, business, and others.

\bibliographystyle{IEEEtran}
\bibliography{paper}

\end{document}